\def\la{\hbox{{\lower -2.5pt\hbox{$<$}}\hskip -8pt\raise
-2.5pt\hbox{$\sim$}}}
\def\ga{\hbox{{\lower -2.5pt\hbox{$>$}}\hskip -8pt\raise
-2.5pt\hbox{$\sim$}}}
\def\ltsima{$\; \buildrel < \over \sim \;$}
\def\simlt{\lower.5ex\hbox{\ltsima}}
\def\gtsima{$\; \buildrel > \over \sim \;$}
\def\simgt{\lower.5ex\hbox{\gtsima}}

\documentstyle[epsfig]{elsart}

\begin{document}
\begin{frontmatter}
\title{Theory of synchrotron radiation: II. Backreaction in 
ensembles of particles}
\author[lngs]{Roberto Aloisio\thanksref{corr1}} 
\address[lngs]{INFN, Laboratori Nazionali del Gran Sasso\\
SS. 17bis, Assergi (AQ), ITALY}
\thanks[corr1]{E-mail: aloisio@lngs.infn.it}
\author[Fermi]{Pasquale Blasi\thanksref{corr}} 
\address[Fermi]{INAF/Istituto Nazionale di Astrofisica\\
Osservatorio Astrofisico di Arcetri\\
Largo E. Fermi, 5 - 50125 Firenze, ITALY}
\thanks[corr]{E-mail: blasi@arcetri.astro.it}

\begin{abstract}
The standard calculations of the synchrotron emission from charged particles
in magnetic fields does not apply when the energy losses of the particles are
so severe that their energy is appreciably degraded during one Larmor rotation.
In these conditions, the intensity and spectrum of the emitted radiation 
depend on the observation time $T_{obs}$: the standard result is recovered
only in the limit $T_{obs}\ll T_{loss}$, where $T_{loss}$ is the time for 
synchrotron losses. In this case the effects of the radiation 
backreaction cannot be detected by the observer.
We calculate the emitted power of the radiation in the most general case,
naturally including both the cases in which the backreaction is relevant and
the standard case, where the usual result is recovered. 
Finally we propose several scenarios of astrophysical interest in which 
the effects of the backreaction cannot and should not be ignored. 
\end{abstract}

\end{frontmatter}

\section{Introduction}

As pointed out in \cite{syn1} (hereafter paper I), some aspects of the 
synchrotron emission of high energy particles did not receive proper 
attention in the literature, mainly because the standard treatment proved 
to be valid over most of the energy range of interest for astrophysical 
applications. Recently, this range of interest changed considerably, mainly 
due to the discovery of radiation processes at ultra-high energies. 
Nevertheless the important corrections to synchrotron emission in this 
regime have been ignored and the standard calculations have been adopted.

In paper I we explored the generalization of the calculations of the 
synchrotron emission from an ensemble of particles radiating coherently.
We considered there the cases of bunches of particles both in the monoenergetic
case and in the case of a spectrum of particles, and for each we established 
the criteria for the synchrotron emission to be coherent. 

In the present paper we explore an effect that becomes relevant at sufficiently
high energies, when the energy lost by a particle during one Larmor gyration
becomes comparable with the energy of the particle itself. We call this
effect backreaction, with may be a slightly improper term.
We find that in this regime there are important corrections to the spectra 
of the emitted radiation in numerous scenarios currently discussed in the 
literature, confirming but extending previous findings of Ref. \cite{nelson},
where the case of a monoenergetic distribution of radiating electrons was
considered. In this paper we also study the situation of an ensemble of
particles with arbitrary energy spectra and find an approximate analytical
solution of the problem of the synchrotron backreaction, that may be useful
to estimate the magnitude of the effect. The formalism used here, introduced 
in paper I and briefly summarized in this paper, allows us to take naturally 
into account possible coherence effects in the synchrotron radiation, 
together with the backreaction. Cases of astrophysical interest in which the
effects of the backreaction are supposed to play an important role will be
discussed.

The paper is structured as follows: in section 2 we describe the backreaction
and the effects that can be expected on simple basis. In section 3 we calculate
the spectrum of the radiation emitted by monoenergetic particles. In section 
4 we generalize the calculation to the case of a spectrum of radiating 
particles and we present analytical approximations that allow us to estimate 
the effects of the backreaction without (or before) being involved in the 
detailed calculations. We present our conclusions in section 5.

\section{The backreaction}

The rate of energy losses of a particle with mass $m$, Lorentz factor 
$\gamma$ and charge $q$ in a magnetic field $B$ can be written in the 
well known form 
\begin{equation}
\frac{dE}{dt} = \frac{2 q^4}{3 m^2 c^3} B^2 \gamma^2 = 
\frac{2 q^4}{3 m^4 c^7} B^2 E^2,
\label{eq:eloss}
\end{equation}
where for simplicity we assumed that the electron moves perpendicular
to the direction of the magnetic field.
The time for a Larmor rotation can be easily calculated as $\tau_B=
2\pi E/q B c$. The (often) hidden assumption of the standard calculations
of synchrotron radiation is that the energy lost by a particle in the
time $\tau_B$ is negligible compared with the particle energy $E$.
This condition is fulfilled when 
\begin{equation}
\frac{\tau_B}{\tau_R} \ll 1 \to E\ll 
\left(\frac{3 m^4 c^8}{4\pi q^3 B}\right)^{1/2}
\label{eq:back}
\end{equation}
where $\tau_R$ is the typical time of energy losses of the particle, defined by
the expression
\begin{equation}
\tau_R^{-1}=\frac{1}{E}\frac{d E}{d t}=\omega_K \gamma=
\frac{2 q^4 B^2}{3 m^4 c^7}E~.
\label{eq:taur}
\end{equation}
For energies larger than the value found in eq. (\ref{eq:back}), the 
standard calculations fail and a new approach is needed. In the present 
paper we introduce this novel approach and show that the difference between 
the predicted spectra and the standard ones may be extremely important and 
in general cannot be ignored. 

Note that from equation (\ref{eq:back}) the region where the backreaction 
becomes important is numerically given by
\begin{equation}
E \gg 1.9\times 10^{4} B_{Gauss}^{-1/2} ~~~\rm GeV
\end{equation}
for electrons, and by
\begin{equation}
E \gg 6.5\times 10^{10} B_{Gauss}^{-1/2} ~~~\rm GeV
\end{equation}
for protons. Here the magnetic field $B_{Gauss}$ is in Gauss. A visual 
picture of the region in the plane $E-B$ where the backreaction is relevant
is provided in fig. 1 for both electrons and protons: the backreaction 
should be taken into account above the solid lines reported in fig. 1.

\begin{figure}[thb]
 \begin{center}
  \mbox{\epsfig{file=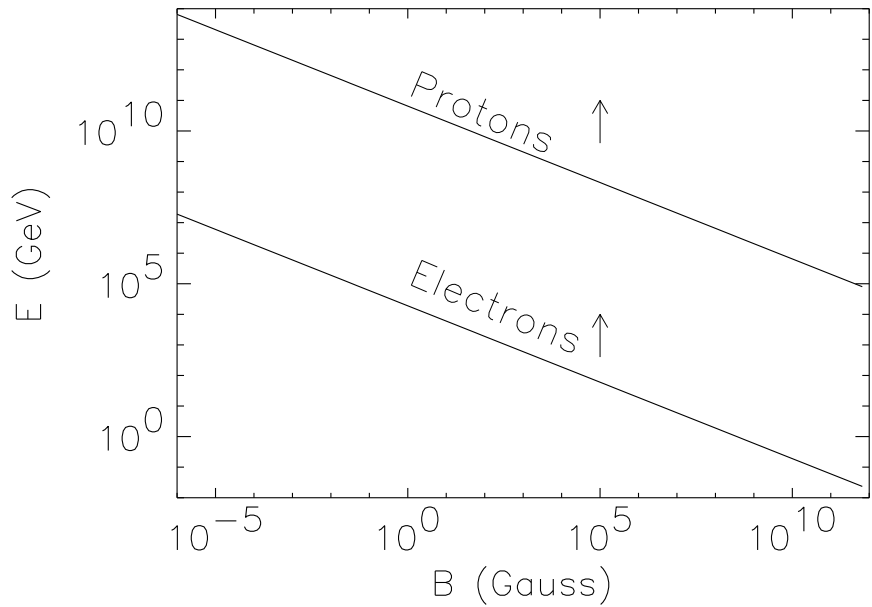,width=13.cm}}
  \caption{\em {Region in plane $E-B$ where the backreaction becomes
relevant. The case of electrons and protons is illustrated.
}}
 \end{center}
\end{figure}

\section{Spectrum of monoenergetic particles}

In this section we compute the spectrum radiated by Z particles,
all having the same initial Lorentz factor $\gamma_0$, as detected by an 
observer during an observation time $T_{obs}$ (this calculation was
also carried out with some differences in the approach, in \cite{nelson}). 
This situation, as shown in \cite{nelson} is equivalent to calculate the 
power radiated by a 
single particle with Lorentz factor $\gamma_0$, as detected by Z
observers during the observation time. Although trivial for the standard
scenario, this is not trivial for the situation of interest here. 

We also demonstrate that for observation times smaller than the time
for appreciable losses, the standard result is recovered.

A nice didactical approach illustrated in \cite{nelson} allows to show that 
the result of the backreaction, that should be derived by the solution of the
Lorentz-Dirac equation, in the end can be determined in the simple way of 
taking into account the time dependence of the Lorentz factor of
the radiating particle.

Based on eq. (\ref{eq:eloss}), we can write the time dependence of the 
Lorentz factor of a particle subject to synchrotron energy losses in the
following form
\begin{equation}
\gamma (t) =\frac{\gamma_0}{1+\omega_R t}
\label{eq:gt}
\end{equation}
where $\omega_R=\omega_K\gamma_0=\tau_R^{-1}$ conveniently defines the time 
scale for the energy losses of a particle. In \cite{nelson} it was shown
that this expression keeps its validity also in case of strong 
backreaction, at least as long as quantum effects remain negligible. These
effects enter the calculation only when the typical energy of the radiated
photon becomes comparable to the energy of the particle, so that quantum 
recoil needs to be accounted for. We will not be involved here 
in these extreme cases, although we believe that some interesting physics
could be learned there. To our knowledge, no paper exists in the literature
treating the problem of the synchrotron emission with both the backreaction 
and the quantum recoil at the same time.

Based on the formalism illustrated in detail in paper I, we associate to
each particle a phase factor $\alpha$. Using the usual relativistic 
beaming condition, implying that significant synchrotron emission is confined
in a narrow cone of aperture $\sim 1/\gamma$, we deduce that a given 
particle illuminates the observer at those retarded times that are related 
to the phase $\alpha$ by the equation
\begin{equation}
t \omega_B(t) - \alpha \approx 0
\label{eq:alpha}
\end{equation}
where $\omega_B(t)=\omega_0 (1+\omega_R t)=
\omega_L\gamma_0^{-1} (1+\omega_R t)$ is the 
(time varying) Larmor frequency of the particle ($\omega_L$ is the usual 
cyclotron frequency). 
From the condition eq. (\ref{eq:alpha}) the i-th particle 
of the ensemble emits its synchrotron burst at the time (we retain here for
obvious reasons only the positive solutions of eq. (\ref{eq:alpha})) 
\begin{equation}
t_i=\frac{1}{2\omega_R} [(1+4\frac{\omega_R}{\omega_0}\alpha_i)^{1/2}-1 ]~.
\label{eq:t}
\end{equation}
From eq. (\ref{eq:t}) it is easy to derive the Lorentz factor of the particle
at the moment of the burst:
\begin{equation}
\gamma_i=\gamma(\alpha_i)=\frac{2\gamma_0}{1+
(1+4\frac{\omega_R}{\omega_0}\alpha_i)^{1/2}}.
\label{eq:gammai}
\end{equation}
Eq. (\ref{eq:gammai}) will be used to rewrite the synchrotron spectrum as 
the spectrum of an ensemble of particles each characterized by a Lorentz 
factor $\gamma_i$ and a phase $\alpha_i$. 

In paper I we have discussed the differences in synchrotron emission 
produced by an ensemble of particles in a coherent and an incoherent 
configuration of phases. In the present paper we will only be concerned 
with the case of incoherent radiation, so that the energy radiated per unit 
frequency and unit solid angle in the directions perpendicular and parallel 
to the magnetic field can be written respectively as follows:
\begin{equation}
\frac{d^2 W_\bot}{d\omega d\Omega}=\frac{q^2}{4\pi c}\sum_{k=1}^{Z}
\left (\frac{\omega}{\omega_L}\right )^2 \frac{\theta_k^4}{\gamma_k^2} 
K_{2/3}^2(\eta_k),
\label{eq:dwpe}
\end{equation}
and
\begin{equation}
\frac{d^2 W_{||}}{d\omega d\Omega}=\frac{q^2}{4\pi c}\sum_{k=1}^{Z}
\left (\frac{\omega}{\omega_L}\right )^2 \theta^2 \theta_k^2
K_{1/3}^2(\eta_k),
\label{eq:dwpa}
\end{equation} 
where we defined the usual quantities \cite{RL,jackson}:
\begin{equation}
\theta_k=(1+\theta^2\gamma_k^2)^{1/2}\qquad \eta_k=
\frac{\omega}{3\omega_L\gamma_k^2}\theta_k^3,
\end{equation}
and $\theta=\pi/2-\theta_z$ with $\theta_z$ the zenith angle of the versor 
$\hat{n}$ that points to the observer.

The quantity that is most interesting from a physical point of view is the 
power (per unit frequency) radiated by the particles. To obtain this quantity,
as we have discussed in paper I, one has to divide the energy per unit 
frequency over the observation time $T_{obs}$. The total power radiated per 
unit frequency by the Z particles in a period $T_{obs}$ 
is\footnote{Here we have used the standard computation of the solid 
angle integration.}
\begin{equation}
\frac{d P}{d \omega}=\frac{\sqrt{3} q^2}{c}\frac{1}{T_{obs}}
\sum_{k=1}^{Z} \frac{\omega}{3\omega_L\gamma_k}\int_{x_k}^{\infty} 
d \xi K_{5/3}(\xi)
\label{eq:pow1}
\end{equation}
where $x_k=\omega/(3\omega_L\gamma_k^2)$.

Consider now the summation over the phases $\alpha_i$, we may consider these
phases homogeneously distributed between $(0,\alpha_M)$, where $\alpha_M$ 
is the maximum value of $\alpha$ such that particles can illuminate the 
observer during the observation time $T_{obs}$ [recall eq. (\ref{eq:t})]:
\begin{equation}
\alpha_M=\omega_L\gamma_0^{-1}T_{obs}(T_{obs}\omega_R + 2)~.
\label{eq:alphaM}
\end{equation} 
Introducing the phase density $\rho(\alpha)=Z/(2\pi)$, we may pass 
from a discrete sum to an integral by the substitution
\begin{equation}
\sum_{k=1}^Z\to \frac{Z}{2\pi}\int_0^{\alpha_M} d\alpha~.
\end{equation}

In conclusion, the power radiated per unit frequency by an ensemble 
of $Z$ particles (all with the same initial Lorentz factor) in the
backreaction regime will be
\begin{equation}
\frac{d P}{d\omega}=\frac{\sqrt{3} q^2}{c}\frac{Z}{2\pi}\frac{1}{T_{obs}}
\int_{0}^{\alpha_M}\frac{\omega}{3\omega_L\gamma(\alpha)}
\int_{x(\alpha)}^{\infty} d\xi K_{5/3}(\xi).
\label{eq:power}
\end{equation}
 
The spectrum found in eq. (\ref{eq:power}) has some general features that 
it is worth to investigate. First, let us check that in the limit 
$T_{obs}\ll \omega_R^{-1}$ the spectrum in eq. (\ref{eq:power}) converges 
to the standard synchrotron spectrum. This condition guarantees that for
small observation times there is no appreciable difference between the 
well known synchrotron spectrum and the predicted one.
It is useful to introduce the variable 
$\Lambda=4\omega_R T_{obs}(\omega_R T_{obs}+2)$ and rewrite the integral
over $\alpha$ as an integral over $y=4\frac{\omega_R}{\omega_0}\alpha$, so
that the power radiated per unit frequency reads
\begin{equation}
\frac{d P}{d\omega}=\frac{\sqrt{3} q^2}{8 c}\frac{Z}{2\pi}
\frac{\omega_0}{\sqrt{1+\frac{1}{4}\Lambda}-1}
\int_0^{\Lambda} d y F(y),
\label{eq:lim}
\end{equation}
where we defined 
$$ F(y)=\gamma(y) x(y)
\int_{x(y)}^{\infty} d\xi K_{5/3}(\xi)~. $$

Now we compute the limit $\Lambda\to 0$ of eq. (\ref{eq:lim}) [note that  
this corresponds to evaluate the limit for $\omega_R T_{obs}\ll 1$]:
\begin{equation}
\lim_{\Lambda \to 0} ~~~\frac{\omega_0}{8\sqrt{1+\frac{1}{4}\Lambda}-1}
\int_{0}^{\Lambda} d y F(y)=\frac{\omega_L}{\gamma_0} F(0).
\end{equation}
Substituting this expression in (\ref{eq:power}) the standard synchrotron
power per unit frequency by an ensemble of $Z$ particles all with 
the same Lorentz factor is readily recovered:
\begin{equation}
\frac{d P}{d\omega}=Z \frac{\sqrt{3} q^2}{c}\frac{\omega_L}{2\pi}
x_0\int_{x_0}^{\infty} d\xi K_{5/3}(\xi),
\label{eq:stdpower}
\end{equation}
where $x_0=\omega/(3\omega_L\gamma_0^2)$.

It is now useful, mainly for practical purposes, to derive analytical 
approximations for the spectrum of the radiation.
We assume first to be in the regime where there are severe modifications due
to the backreaction, that is $(\omega_R T_{obs}\ge 1)$. 
To perform the calculation we will adopt the following rough
approximation of the Bessel function:
\begin{displaymath}
x\int_{x}^{\infty}d\xi K_{5/3}(\xi)=2^{2/3}\Gamma\left ( \frac{2}{3} \right )
\left\{\begin{array}{cc}
x^{1/3} & \qquad 0\le x\le 1 \\
0 & \qquad x> 1 
\end{array}\right.
\end{displaymath}
Although certainly not sophisticated, this approximation allows to treat
analytically expressions that would otherwise be only of numerical access.
The condition $x(\alpha)\le 1$,
using equation (\ref{eq:gammai}) and the definition of $x$, implies that
\begin{equation}
\alpha\le \gamma_0 \frac{\omega_0}{\omega_R}
\left (\frac{\omega}{3\omega_L} \right)^{-1}\left [\gamma_0-
\left (\frac{\omega}{3\omega_L}\right )^{1/2} \right ]=\alpha_0
\label{eq:alpha0}
\end{equation}
Therefore, to use the approximation of the Bessel function, we have to perform
the integration over $\alpha$ between $(0,\tilde{\alpha})$ where 
$\tilde{\alpha}=\min\{\alpha_M,\alpha_0\}$. The condition 
$\alpha_0 = \alpha_M$ determines the frequency where there is a change
in the slope of the spectrum of the emitted radiation. This identifies
two frequency regimes, the low frequency one ($\tilde{\alpha}=\alpha_M$)
and the high frequency one ($\tilde{\alpha}=\alpha_0$). 
The separation between the two regimes occurs at 
\begin{equation}
\omega = 3\omega_L\gamma_0^2 Y(T_{obs})
\label{eq:lowfre}
\end{equation}
where 
$$
Y(T_{obs})=\left [ \frac{\sqrt{1+4(\omega_R T_{obs})^2+8(\omega_R T_{obs})}-1}
{2\omega_R T_{obs}(\omega_R T_{obs}+2)} \right ]^2 
$$
and, using the condition of strong backreaction $\omega_R T_{obs}\ge 1$, 
we may approximate
\begin{equation}
Y(T_{obs})\simeq \frac{1}{(\omega_R T_{obs})^2}~.
\label{eq:Y}
\end{equation}

In the high frequency regime $3\omega_L\gamma_0^2 Y(T_{obs})<\omega<
3\omega_L\gamma_0^2$ the integration over $\alpha$ has to be performed 
in the range $(0,\alpha_0)$.

Within the approximations adopted here, we get the following expression
for the radiated power:
\begin{itemize}

\item{Low frequency \\
$\omega<3\omega_L\gamma_0^2 Y(T_{obs})$ \\

\begin{equation}
\frac{d P}{d\omega}=\frac{\sqrt{3} q^2}{c} \frac{Z}{2\pi} \frac{3}{5}
2^{2/3}\Gamma \left (\frac{2}{3} \right ) 
\omega_L(\omega_R T_{obs})^{2/3} 
\left (\frac{\omega}{3\omega_L\gamma_0^2}\right )^{1/3}
\label{eq:powerlow}
\end{equation}
}

\item{High frequency \\
$3\omega_L\gamma_0^2 Y(T_{obs})<\omega<3\omega_L\gamma_0^2$ \\ 

\begin{equation}
\frac{d P}{d\omega}=\frac{\sqrt{3} q^2}{c} \frac{Z}{2\pi} \frac{3}{5}
2^{2/3} \Gamma \left (\frac{2}{3} \right ) 
\frac{\omega_L}{\omega_R T_{obs}} 
\left (\frac{\omega}{3\omega_L\gamma_0^2}\right )^{-1/2}
\label{eq:powerhigh}
\end{equation}
}

\end{itemize} 

Eqs. (\ref{eq:powerlow},\ref{eq:powerhigh}) are very interesting and
allow a simple interpretation of the backreaction regime. 
In the low frequency regime the spectrum of the synchrotron radiation 
has the same slope as in the standard case ($\omega^{1/3}$); on the other 
hand, moving to high frequency, the spectrum has a new power law behaviour 
of the type $\omega^{-1/2}$. Moreover, in the backreaction regime, the 
maximum of the spectrum is located at a frequency that depends on the 
observation time according to the following expression:
\begin{equation}
\omega_{cr}(T_{obs})=3\omega_L\gamma_0^2 Y(T_{obs})\simeq 
\frac{3\omega_L\gamma_0^2}{(\omega_R T_{obs})^2}=
\frac{3\omega_L}{(\omega_K T_{obs})^2}~.
\label{eq:ommax2}
\end{equation}
In other words the maximum gradually moves toward lower frequencies when 
the observation time is increased and its position does not depend on 
$\gamma_0$: the integrated power becomes gradually richer in its
low frequency component.

We plotted our results for the radiation spectra in fig. 2 for three 
cases as indicated: 1) standard case; 2) $\Lambda=10^4$ and 3) 
$\Lambda=10^6$. Larger values of $\Lambda$ correspond to higher levels
of backreaction, so that the spectra are gradually peaked at lower
frequencies when the observation time increases. For the two cases of
backreaction, we also plot our analytical approximations for the low
and high frequency regimes. At low frequency the agreement with the 
detailed calculation is excellent. At higher frequencies clearly the 
rough approximation adopted for the Bessel functions gives a poorer
but still acceptable agreement, very useful for practical estimates.
In particular the slopes predicted by the analytical calculations reproduce
very well the ones obtained in the detailed calculations. 

Note that the plot in fig. 2 is made in such a way that can be applied 
equally well to electrons and protons as radiating particles and for
any value of the magnetic field. All these parameters in fact enter the
definition of $\omega_L$. 

\begin{figure}[thb]
 \begin{center}
  \mbox{\epsfig{file=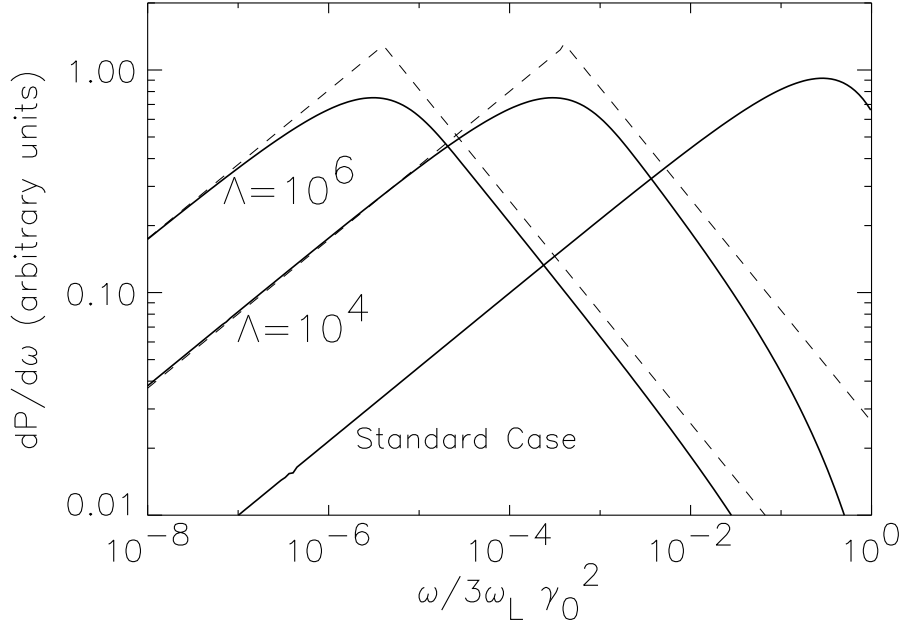,width=13.cm}}
  \caption{\em {Power spectrum of the radiation emitted by monoenergetic
particles for the standard case (no backreaction) and for the cases $\Lambda=
10^4$ and $\Lambda=10^6$ (The dashed lines represent the results of the 
analytical approximation illustrated in the text).
}}
 \end{center}
\end{figure}

\section{Synchrotron radiation from particles with a power law spectrum}

We assume here to have a spectrum of charged particles in the form
\begin{equation}
N(\gamma_0)=N_0\gamma_0^{-p}
\label{eq:distr}
\end{equation}
where $p>1$ is a spectral index. For most astrophysical applications this is 
the relevant case, therefore we explore here in detail what are our 
predictions for the spectrum of the synchrotron emission.
Note that $N(\gamma_0)$ is here what is usually called the equilibrium 
spectrum of the radiating particles, as derived from a transport equation,
usually including energy losses. This spectrum is typically steeper than 
the injection spectrum. The equilibrium spectrum is here taken to be
time independent, meaning that there is a continuous replenishment of
the particles at all energies, requiring the source to be active for the
all duration of the phenomenon. The case of sources bursting on time
scales much shorter than the observation time will not be considered
here, but can be easily recovered by generalizing the discussion below.

Starting from eq. (\ref{eq:power}) we can generalize the calculations 
of the previous section simply by the substitution $Z\to N(\gamma_0)$ and 
introducing an integration over the energy spectrum of the particles. The
power radiated will be
\begin{equation}
\frac{d P}{d\omega}=\frac{\sqrt{3} q^2}{c}\frac{1}{2\pi}\frac{1}{T_{obs}}
\int_{\gamma_{min}}^{\gamma_{max}}
N_0\gamma_0^{-p}
\int_{0}^{\alpha_M}\frac{\omega}{3\omega_L\gamma(\alpha)}
\int_{x(\alpha)}^{\infty} d\xi K_{5/3}(\xi)
\label{eq:powerdist}
\end{equation}
where $\gamma_{min}$ and $\gamma_{max}$ are the minimum and maximum values 
allowed for the Lorentz factor, fixed by the particular physical system under 
consideration.

Using the same approximation of the Bessel function introduced in the 
previous section we will work out from equation (\ref{eq:powerdist}) some
interesting analytical results. Let us first assume that, fixing 
$T_{obs}$, for any $\gamma_0$ inside the interval 
$(\gamma_{min},\gamma_{max})$ the system is in the backreaction 
regime $\omega_R T_{obs} \ge 1$.
Using the high and low frequency results of the previous section
(cfr. eqs. (\ref{eq:powerhigh}) and (\ref{eq:powerlow})
we may calculate the integral over $\gamma_0$. 

Let us start from the high frequency regime, 
$3\omega_L\gamma_0^2 Y(T_{obs})<\omega<3\omega_L\gamma_0^2$. 
In this case, fixing the frequency $\omega$ from the condition 
$\omega<3\omega_L\gamma_0^2$ one obtains
$$ \gamma_0>\left (\frac{\omega}{3\omega_L}\right )^{1/2}=\tilde{\gamma} $$
therefore, for consistency, the integral over $\gamma_0$ has to 
be calculated between $\tilde{\gamma}$ and $\gamma_{max}$. In this case 
(recalling eq. (\ref{eq:powerhigh})) the power radiated per unit 
frequency is
\begin{equation}
\frac{d P}{d\omega}=A \frac{\omega_L}{\omega_K T_{obs}}
\left (\frac{\omega}{3\omega_L} \right )^{-1/2}
\int_{\tilde{\gamma}}^{\gamma_{max}} d\gamma_0 \gamma_0^{-p}
\label{eq:po}
\end{equation} 
where $A$ is a numerical factor defined as
$$A=\frac{\sqrt{3} q^2}{2\pi c}\frac{3}{5} 2^{2/3}
\Gamma\left (\frac{2}{3} \right )N_0~.$$ 
Assuming that 
$\gamma_{max}\gg \tilde{\gamma}$ we may consider the integral 
in (\ref{eq:po}) from 
$(\tilde{\gamma},\infty)$ obtaining the high frequency power emitted
per unit frequency
\begin{equation}
\left (\frac{d P}{d\omega}\right )_{high}^{back}=A \omega_L
\frac{1}{\omega_K T_{obs}}
\frac{1}{p-1}
\left (\frac{\omega}{3\omega_L} \right )^{-p/2}.
\label{eq:powhidis}
\end{equation}  

Let us now evaluate the power emitted in the low frequency regime
$\omega<3\omega_L Y(T_{obs})$. Recalling that we are considering the case in
which the backreaction is always important by construction, 
for any $\gamma_0\in(\gamma_{min},\gamma_{max})$, one 
can approximate $Y(T_{obs})\simeq 1/(\omega_R T_{obs})^2$ (cfr. eq. 
(\ref{eq:Y})) and therefore
$$\omega<\frac{3\omega_L}{(\omega_K T_{obs})^2}~. $$
In this case there is no condition on the lower extreme in the integral over
$\gamma_0$, so that, assuming $\gamma_{min}\ll \gamma_{max}$, the low 
frequency power emitted per unit frequency can be written as 
\begin{equation}
\left (\frac{d P}{d\omega}\right )_{low}^{back}
=A \omega_L(\omega_K T_{obs})^{2/3}
\frac{\gamma_{min}^{-(p-1)}}{p-1}
\left (\frac{\omega}{3\omega_L} \right )^{1/3}.
\label{eq:powlodis}
\end{equation}

It is instructive to study the frequency corresponding to the 
change from the low to the high frequency regime. As we have pointed 
out before [eq. (\ref{eq:Y})], this frequency depends on the observation 
time according to the following expression:
\begin{equation}
\omega_{cr}=3\omega_L \gamma_0^2 Y(T_{obs})\simeq 
\frac{3\omega_L}{(\omega_K T_{obs})^2}~.
\label{eq:omcri}
\end{equation}
This expression should be compared with the critical frequency defined within 
the standard theory of synchrotron emission for the particles
with the minimum Lorentz factor, $\omega_{syn}=3\omega_L\gamma_{min}^2$.
Since we assumed that, for any $\gamma_0$, we are in the backreaction regime 
($\omega_K\gamma_{min} T_{obs}\ge 1$), we have 
$$\omega_{cr}\ll \omega_{syn}$$
therefore, the change in the behaviour of the power radiated per unit 
frequency occurs in the low frequency regime.

We are now ready to discuss a more realistic situation, for which
there is a value $\gamma_c$ with $\gamma_{min}<\gamma_c<\gamma_{max}$ 
that divides, for any fixed observation time, the standard regime from 
the backreaction one. This value of $\gamma_c$ is simply given by
$$\gamma_c=\frac{1}{\omega_K T_{obs}}~.$$

In the situation for which $\gamma_c\in(\gamma_{min},\gamma_{max})$ we can 
divide the spectrum in two regions: one for frequencies
$\omega<3\omega_L\gamma_c^2$, in which backreaction is not important, and the 
other, for $\omega>3\omega_L\gamma_c^2$, in which the backreaction effects
need to be included. 

In the situation in which $\gamma_c\in(\gamma_{min},\gamma_{max})$
the spectrum radiated by particles is
\begin{itemize}
\item{At high frequencies, $\omega>3\omega_L\gamma_c^2$: \\

\begin{equation}
\frac{d P}{d\omega}=\left (\frac{d P}{d\omega}\right )_{high}^{back}
\label{eq:highh}
\end{equation}
}
\item{At low frequencies, $\omega<3\omega_L\gamma_c^2$: \\

\begin{equation}
\frac{d P}{d\omega}=\left (\frac{d P}{d\omega}\right )^{syn}+  
\left (\frac{d P}{d\omega}\right )_{low}^{back}
\label{eq:loww}
\end{equation}
}
where the first term is the standard synchrotron spectrum.
\end{itemize}
 
The standard synchrotron spectrum (integrated between 
$\gamma_{min}$ and $\gamma_c$) has the following form:
$$
\left (\frac{d P}{d\omega}\right )^{syn}
=N_0 \frac{\sqrt{3} q^2}{c}
\frac{\omega_L}{2\pi}2^{2/3}\Gamma\left(\frac{2}{3}\right)
\frac{3}{3p-1} 
$$
\begin{equation}
\left[
\left (\frac{\omega}{3\omega_L} \right )^{-\frac{p-1}{2}}-
\left (\frac{\omega}{3\omega_L} \right )^{\frac{1}{3}}\gamma_c^{-p+\frac{1}{3}}
\right]
\label{eq:std}
\end{equation}  
while the backreaction dominated spectrum is described by eqs.
(\ref{eq:powhidis}) and (\ref{eq:powlodis}) with $\gamma_{min}=\gamma_c$. 
Let us now compare the two components of the low frequency spectrum. For 
this purpose we evaluate explicitly eq. (\ref{eq:loww})
$$
\left (\frac{d P}{d\omega}\right )_{low}^{back}+ 
\left (\frac{d P}{d\omega}\right )^{syn}=$$
\begin{equation}
C\cdot\left\{ \frac{3}{3p-1}\left (
\frac{\omega}{3\omega_L}\right )^{-\frac{p-1}{2}}
+(\omega_K T_{obs})^{p-\frac{1}{3}}
\left (\frac{\omega}{3\omega_L}\right )^{\frac{1}{3}} 
\left [\frac{3}{5}\frac{1}{p-1}-\frac{3}{3p-1}\right ]\right \}
\label{eq:conf}
\end{equation}
where $C$ is a numerical factor 
$$C=N_0 \frac{\sqrt{3} q^2}{c}\frac{\omega_L}{2\pi} 2^{2/3}
\Gamma\left (\frac{2}{3}\right )$$
and we have used the relation $\gamma_c=1/(\omega_K T_{obs})$.

It is easy to see that at frequencies $\omega\ll \omega_{cr}$ the low 
frequency spectrum is well described by the standard synchrotron emission 
(i.e. the second term in (\ref{eq:conf}) is always negligible). Moreover,
this is also rigorously true in the special case $p=2$, as it is 
easy to show by explicitly calculating the second term in eq. 
(\ref{eq:conf}).

\begin{figure}[thb]
 \begin{center}
  \mbox{\epsfig{file=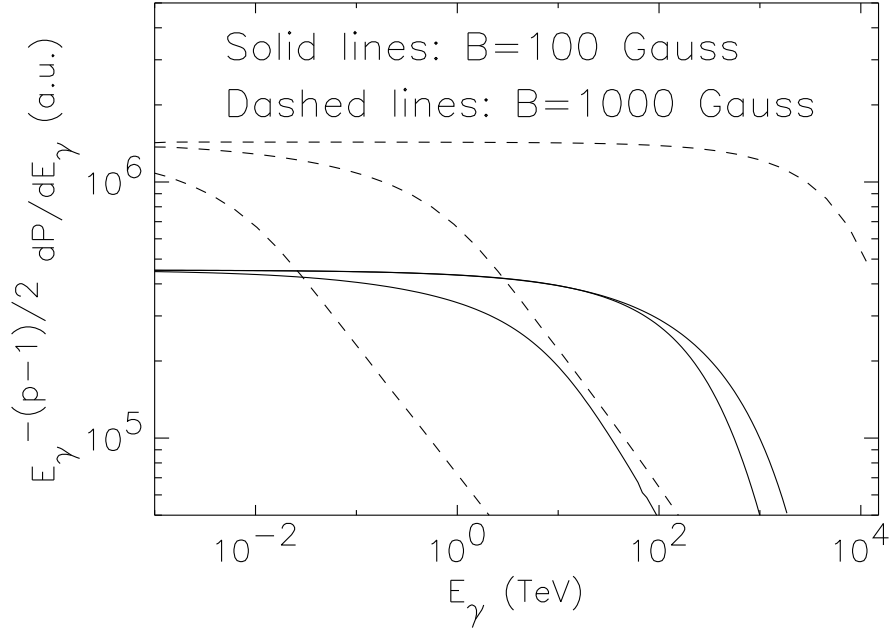,width=13.cm}}
  \caption{\em {Power spectrum of the radiation emitted by protons
with a power law spectrum $N(\gamma_0)=N_0 \gamma_0^{-p}$ for the standard 
case ($T_{obs}\to 0$) and for $T_{obs}=10^3~s$ and $T_{obs}=10^4~s$ 
(moving leftward). The two sets of curves are for $B=100$ Gauss (solid lines)
and $B=1000$ Gauss (dashed lines).
}}
 \end{center}
\end{figure}

In conclusion, in the most interesting case in which radiation reaction is
efficient only for $\gamma>\gamma_c>\gamma_{min}$, the power radiated by 
the ensemble of particles with a fixed (power law) distribution of Lorentz
factors is described at low frequency $(\omega<3\omega_L\gamma_c^2)$ by the 
standard synchrotron spectrum (\ref{eq:std}) and at high frequency
($\omega>3\omega_L\gamma_c^2$) by the backreaction modified spectrum in eq.
(\ref{eq:powhidis}).

In fig. 3 we plot for illustrative purposes the results of a specific case:
high energy protons with a spectrum $\propto E^{-2}$ and high energy cutoff
at $10^{21}$ eV radiating by synchrotron emission in magnetic fields of order
$100$ Gauss (solid lines) and $1000$ Gauss (dashed lines). The curves are
obtained for three choices of the observation time $T_{obs}$. The rightmost 
curves in both cases correspond to the standard synchrotron emission (small
observation time) while moving leftward we plot the cases $T_{obs}=10^3~s$
and $T_{obs}=10^4~s$. The effect of the increasing backreaction is clear from
these curves. This specific case has been chosen because the situation is similar
to the one proposed in \cite{aharonian,protheroe} to explain the TeV emission
from BL Lac as synchrotron emission of ultra high energy protons (in these 
papers the standard synchrotron emission was adopted). 

\section{Conclusions}

We calculated the spectrum of the synchrotron emission from a system 
of charged particles in the most general case in which the observation 
time is arbitrary compared to the time for the energy losses of the 
particles. We first calculated the effect for the case of a  
particle with Lorentz factor $\gamma_0$.

Our findings on the monoenergetic case can be summarized as follows:

{\it i)} There is a critical frequency $\omega_{cr}$ such that the spectrum 
of the radiation at $\omega\le \omega_{cr}$ is the usual synchrotron spectrum
$\propto \omega^{1/3}$, while for $\omega\ge \omega_{cr}$ the backreaction 
affects the spectrum changing it to $\propto \omega^{-1/2}$.

{\it ii)} The critical frequency $\omega_{cr}$ depends on the observation time
but it turns out to be independent of the Lorentz factor of the radiating
particles $\gamma_0$. The dependence on these parameters is as found in 
eq. (\ref{eq:ommax2}), so that increasing the observation time, power
is moved to gradually lower frequencies. 

{\it iii)} The standard limit is recovered when the observation time is
much smaller than the time for losses $\omega_R T_{obs}\ll 1$.

The more realistic case investigated in this paper is that of a power
law spectrum of particles $N(\gamma_0)\propto \gamma_0^{-p}$. In general 
what happens is that there may be a critical Lorentz factor $\gamma_c$, 
such that particles with $\gamma_0\ge\gamma_c$ are affected by the 
backreaction while the particles with $\gamma_0\le \gamma_c$ behave in 
the standard way. The spectrum of the radiation in this case is a 
superposition of different components. Our findings can be summarized 
as follows:

{\it a)} the spectrum of the radiation can be divided into a low frequency
one and a high frequency one, with the separation occurring at the frequency
$\omega_{cr}=3\omega_L/(\omega_K T_{obs})^2$. 

{\it b)} the particles with Lorentz factor $\gamma_0\le\gamma_c$ radiate
the standard synchrotron radiation whose spectrum is 
$\propto \omega^{-(p-1)/2}$.

{\it c)} the particles with $\gamma_0\ge\gamma_c$ do radiate in regime of
backreaction and affect both the low frequency and the high frequency regime.

{\it d)} the low frequency spectrum radiated by particles with 
$\gamma_0\ge\gamma_c$ is $\propto \omega^{1/3}$ for $\omega\ll \omega_{cr}$,
but it is always dominated, in the same frequency range, by the standard
synchrotron radiation.

{\it e)} the high frequency radiation radiated by particles with 
$\gamma_0\ge\gamma_c$ is $\propto \omega^{-p/2}$. Therefore it represents
a suppression of the radiation compared to the case of standard synchrotron
radiation. Note that increasing the observation time, while the slopes at
low and high frequency remain unchanged, the boundary between the two regimes
moves toward lower frequencies, so that, as a consequence, the height of 
the spectrum at high frequencies becomes increasingly lower. 

{\it f)} At fixed observation time (which is obviously decided by the
observer) the critical frequency only depends on the magnetic field in
the production region. This is a very important point: in the standard
synchrotron emission, it is in general not possible to extract the 
magnetic field from a measurement of the synchrotron flux because there
is degeneracy between the number of radiating particles and the strength 
of the magnetic field. In the backreaction case however, the position of 
the change in slope uniquely defines the magnetic field, so that the
measurement can in principle be used to directly infer the strength of
the magnetic field.

The crucial question is whether there are situations in which the conditions
for the backreaction to be relevant are fulfilled. The answer can be found in
fig. 1, where we plotted the regions of interest for both electrons and 
protons in the plane $B-E$. At each magnetic field there corresponds a range
of energies for which the backreaction is important. We immediately see that
for conditions typical in the Galaxy, $B\sim 1\mu G$ only electrons with 
$E\simgt 10^7$ GeV feel the effects of the backreaction. On the other hand 
in magnetic fields which are typical of neutron stars, $B\sim 10^{10}-10^{12}$
Gauss, electrons with energies in excess of MeV-GeV already need to be 
accounted for in the frame of a backreaction approach. In the same environment,
protons with energies larger than $10^5-10^6$ GeV also radiate in a regime 
in which backreaction is important.

Some applications of the calculations and results illustrated in this 
paper will be presented in forthcoming papers. They include: the
synchrotron emission from ultra-high energy electron-positron pairs 
generated as a result of the decay of super-heavy relics \cite{bere}
in the Galaxy {\cite{blasi} or in the decay of the $Z^0$ resonance produced 
in $\nu\bar\nu$ annihilation (Z-burst model \cite{weiler,fargion}). 
In this case we have magnetic fields $B\sim 1\mu G$ and
energies in excess of $10^{10}$ GeV, so that the synchrotron radiation is
strongly affected by the backreaction. 

A more {\it conventional} application concerns the TeV gamma ray emission
from BL Lac objects. In \cite{aharonian,protheroe} a proton synchrotron 
model was proposed in which the TeV emission is the result of the synchrotron
emission of ultra high energy protons ($E\sim 10^{19}$ eV) in a magnetic field
of order $\sim 100$ Gauss. We estimate that with these parameters and for 
observation times of the order of the duration of the observed flares (a few
hours) the backreaction affects visibly the TeV gamma ray spectra 
\cite{AloBla}.

{\bf Aknowledgments} We are very grateful to F. Pacini and A. Olinto 
for many useful discussions and to M. Salvati and T. Stanev for a critical 
reading of the manuscript. We are also grateful to the anonymous referee
for the interesting remarks on the paper.

\end{document}